\documentclass[conference]{IEEEtran}
 
\usepackage{cite}

\usepackage{gensymb}
\usepackage{soul}

\usepackage{booktabs}
\usepackage{array}

\ifCLASSINFOpdf
\usepackage[pdftex]{graphicx}
\usepackage[dvipsnames]{xcolor}
   
\else
   
\fi

\usepackage[cmex10]{amsmath}

\usepackage{algorithmic}

\usepackage{mdwmath}
\usepackage{mdwtab}
 
\usepackage[justification=centering]{caption}

\usepackage[tight,footnotesize]{subfigure}

\usepackage{placeins}

\begin{document}
%
\title{Generalized Circuit Averaging Technique for Two Switch DC-DC Converters} 

\author{\IEEEauthorblockN{Sumukh Surya$^{\gamma}$, \textit{Intern},  Janamejaya Channegowda$^\delta$, \textit{Member, IEEE}, Kali Naraharisetti$^{\beta}$ \\
$^{\gamma}$e-Powertrain, KPIT, Bangalore, India, sumukhsurya@gmail.com \\
$^\delta$Ramaiah Institute of Technology, Bangalore, bcjanmay.edu@gmail.com}
$^\beta$Infineon Technologies, swaraj.kali@gmail.com}
\maketitle

\begin{abstract} 
Cuk and SEPIC are some of the important DC-DC converters used for charging batteries. In this paper, a generalized circuit averaging technique is employed for Cuk and SEPIC converters. The derived equations are used to obtain the frequency response of open loop transfer function. The ratio of perturbed output voltage to duty cycle ($G_{vd}$) is simulated using LTSpice software package. The derived  averaged models of the converters aids in faster and simpler simulation. The behavior of the converters in CCM and DCM was also simulated. The derived expressions can be generalized to power converters with two switches.
\end{abstract}

\IEEEpeerreviewmaketitle

\textbf{Keywords:}  \textbf{Circuit Averaging, Cuk, DCM, CCM Instability, LTSpice, SEPIC}

 \section{Introduction} 
DC-DC converters have gained popularity due to the emergence of Electric Vehicles (EVs). DC-DC converters can be classified into isolated and non-isolated topologies. Among various non-isolated converters, fourth order converters like Cuk and SEPIC have been given prominence as they provide advantages such as non-inverted output voltage and ability to operate from an input source which has a value greater or lesser than output voltage. 

Determining open loop transfer function for such converters plays an important role as they provide useful info to asses converter stability and help improve controller design. In literature, several attempts have been made in examining different approaches for obtaining the transfer function. 
In [1], three different approaches such as: a) Small signal model, b) Circuit Averaging and c) State Space averaging for DC-DC converters in CCM and DCM was introduced. 

It was shown that the losses in the converters are primarily contributed by switching and not by conduction. In [2], a SEPIC operating in DCM was selected to drive a Light-Emitting Diode (LED) for constant voltage application. An average and a switching model was developed, modeled in MATLAB / Simulink and validated against the experimental results. The transfer functions $G_{vd}$ and $G_{vg}$ (Output voltage to input voltage) were derived. It was shown that the SEPIC provided lower input current harmonics.  

In [3], SEPIC was modelled for DCM operation by using State Space averaging technique, implemented using MATLAB and LTSpice simulation tools. It was shown that the Bode plots obtained from these tools closely matched experimental results at frequencies below 10 kHz. At higher frequencies, the simulation plots diverged from experimental results due to reduced order matrix.

In [4], an ideal SEPIC and Cuk operating in DCM are selected and used for Power Factor Correction (PFC). The advantages of the converters is discussed in detail. The input to the converters is supplied by a single phase rectifier. The open loop transfer function obtained using the small signal model are validated against the hardware results and they were found to be closely correlated.  

In [5], concept of circuit averaging for converters like Buck, Boost and Buck-Boost in DCM was discussed. It was shown that the input and output ports of such converters behave like a resistive and power sink respectively. 

In [6], an averaged model in LTSpice was developed for ideal Buck and Boost operating in CCM was constructed using CCM block available in LTSpice software package. 

In [7], a mathematical model for Cuk converter operating in CCM was derived and modelled using Simulink. The importance of step size while capturing the transients was shown. \\
In this paper, DCM analysis for practical converters, SEPIC and Cuk are carried out using Circuit Averaging using LTSpice simulation tool. It was found that the cause of discontinuity in ideal and non-ideal converters was due to the sum of inductor currents ($i_{L1}$ + $i_{L2}$) being zero. CCM-DCM block in LTSpice was used which solves for the various currents and voltages independent of the operation of the converter.

\section{Circuit Averaging for an Ideal SEPIC}

\begin{figure}[htbp!]
\centering
\includegraphics[scale=1.5]{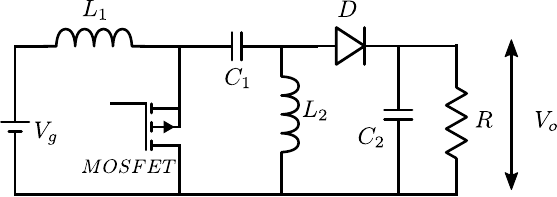}
\caption{Circuit diagram of SEPIC}
\label{fig:SEPIC}
\end{figure}
 
Fig. 1 shows an ideal SEPIC with two switches MOSFET and diode. The voltage across the MOSFET and diode are named as $V_1$ and $V_2$ respectively. Similarly, the current in MOSFET and diode are named as $I_1$ and $I_2$ respectively. \\
Circuit averaging of any converter involves three major steps:

\begin{enumerate}
\item Separate the switch network from the converter and define the ports
\item Sketch the waveform of the switch current and voltage waveforms followed by averaging
\item Simplify equations and draw the equivalent switch network
\end{enumerate}
 
\begin{figure}[htbp!]
\centering
\includegraphics[scale=0.85]{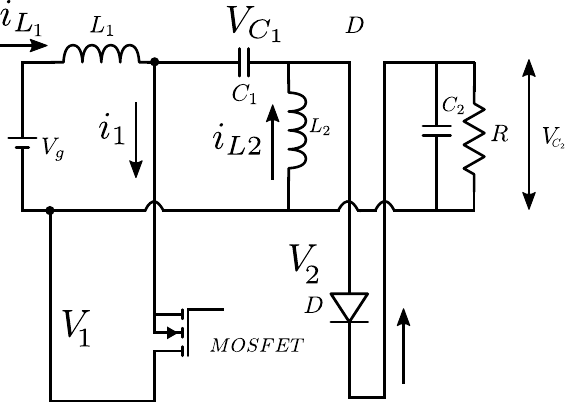}
\caption{Separating the Switches}
\label{fig:Separate}
\end{figure}
\begin{figure}[htbp!]
\centering
\includegraphics[scale=1.5]{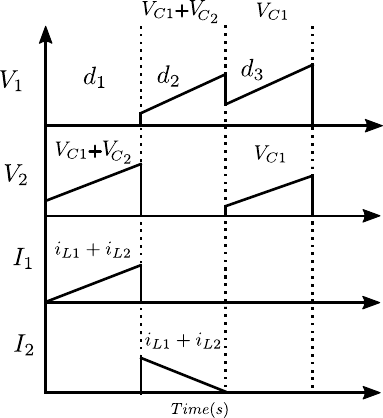}
\caption{Averaged Switch Voltages and Currents}
\label{fig:Average}
\end{figure}
 
The DCM operation occurs due to the unidirectional flow of current in the switch (Diode). Hence, the sum of inductor currents ($i_{L1}$ +$i_{L2}$) contribute to the DCM operation in Cuk converter and SEPIC.\\
Averaging $I_1$ (Input Port)
\begin{align} 
<I_1> &=  \frac{1}{2T_s} * (D_1T_s) *(i_{peak_{L1}}+i_{peak_{L2}}) \\
i_{peakL1}+i_{peakL2}&= \frac{(V_1D_1T_s)}{L}\\
Substituting \hspace*{0.2cm} (2) \hspace*{0.2cm} in \hspace*{0.2cm} (1), \\
I_1 &= \frac{(D^2_1V_1T_s)}{2L}\\
where \hspace*{0.2cm} L&=\frac{(L_1L_2)}{L_1+L_2}\\
Since, \hspace*{0.2cm} T_s&=\frac{1}{f_s}\\
\frac{V_1}{I_1} &= \frac{2Lf_s}{D^2_1} \\
Where \hspace*{0.2cm} R_e&=\frac{2Lf_s}{D^2_1} \\
And \hspace*{0.2cm} D_1&=D
\end{align}
Hence, the input port behaves like a loss free resistor, though physically no resistor exits. Averaging the current waveform at the output port,we obtain:
\begin{align} 
<I_2> &=  \frac{1}{2T_s} * (D_2T_s) *(i_{peak_{L1}}+i_{peak_{L2}}) \\
i_{peakL1}+i_{peakL2} &= \frac{(V_2D_2T_s)}{L} \\
D_1(V_1)&=D_2(V_2)\
\end{align} 
As observed from Fig. 3, inductors charge from zero and reach the peak value in $D_1$$T_s$. However, the same currents reach zero in $D_2$$T_s$ interval.
Substituting (11) and (12) in (10) we get
\begin{align}
<I_2> &=  \frac{1}{2L} * (D_2^2V_2T_s)  \\
D_2 &=  \frac{D_1V_1}{V_2}
\end{align}
\begin{align}
<I_2> &=  \frac{V_1^2D_1^2}{2V_2Lf_s}  \\
I_2V_2 &= \frac{V_1^2}{R_e}
\end{align} 
\begin{figure}[htbp!]
\centering
\includegraphics[scale=1.5]{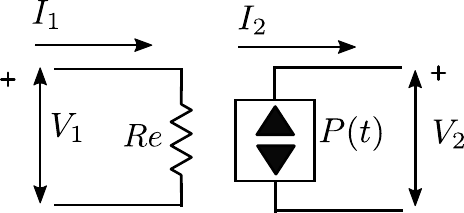}
\caption{Equivalent Circuit for the switch network}
\label{fig:Average}
\end{figure}

\section{Circuit Averaging for a NON-Ideal SEPIC}
\begin{figure}[htbp!]
\centering
\includegraphics[scale=0.85]{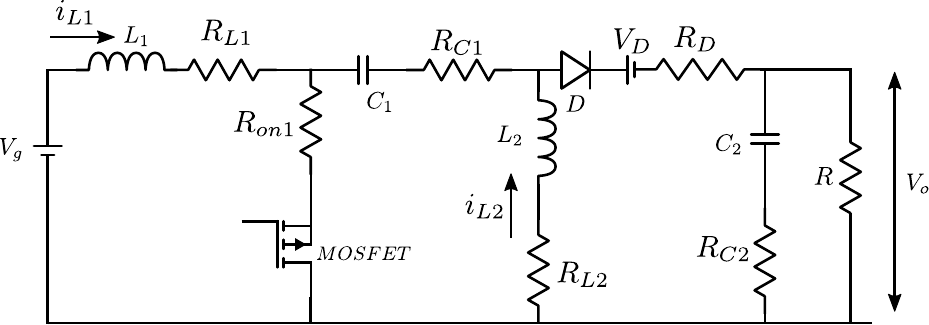}
\caption{Circuit Diagram of a Non-Ideal SEPIC}
\label{fig:Non-Ideal SEPIC}
\end{figure}
Fig. 5 shows a non-ideal SEPIC. The MOSFET and the diode have to be separated from the circuit as shown in Fig. 2.\\
Fig. 6 shows the separation of the switches from the circuit with $V_0$ = $V_{c2}$
\begin{figure}[htbp!]
\centering
\includegraphics[scale=1.2]{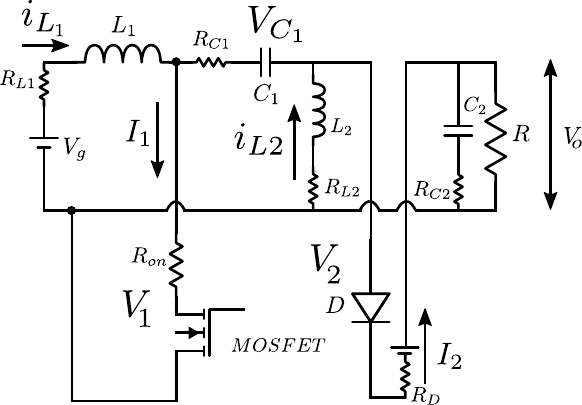}
\caption{Swicthes Separated}
\label{fig:Swicthes separated}
\end{figure}

\begin{align} 
<V_1> &=((i_{L1}+i_{L2}R_{on1}D_1+((V_{c1}+V_{c2}+V 
\end{align}
\begin{align*} 
+R_d(i_{L1} + i_{L2})D_2+V_{c1}D_3 
\end{align*}
\begin{align} 
<V_2> &=((V_{c1}+V_{c2})-(i_{L1}+i_{L2})R_{on1})D_1
\end{align}
\begin{align*} 
D_2(V_d+R_d(i_{L1}+i_{L2}))+D_3(V{0}) 
\end{align*}
\begin{align} 
<I_1> &=D_1*(i_{L1}+i_{L2}) 
\end{align}
\begin{align} 
<I_1> &=D_2*(i_{L1}+i_{L2}) 
\end{align}
 
%

From (19) and (20), it can be observed that the governing equation to describe DCM in a non-ideal SEPIC is similar to that of (15) and (16)
Hence, the equivalent switch network is similar to Fig. 4
\section{Circuit Averaging for an Ideal Cuk}
Fig. 7 shows an ideal Cuk converter operating in DCM. Fig. 8 shows the MOSFET and Diode separated from the converter.
\begin{figure}[htbp!]
\centering
\includegraphics[scale=0.85]{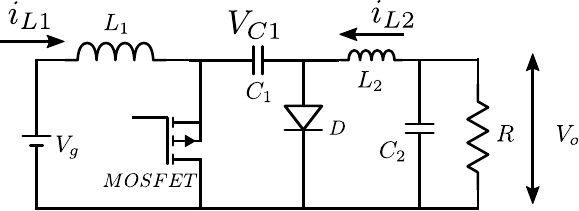}
\caption{Ideal Cuk converter}
\label{fig:Ideal Converter}
\end{figure}
\begin{figure}[htbp!]
\centering
\includegraphics[scale=0.85]{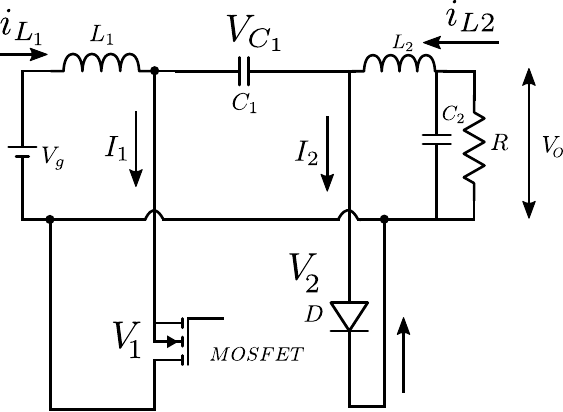}
\caption{Switches Separated}
\label{fig:Switches Separated}
\end{figure}

Fig. 9 shows the waveforms of switch voltages and currents at $D_1T_s$, $D_2T_s$ and $D_3T_s$ intervals.

\begin{figure}[htbp!]
\centering
\includegraphics[scale=1]{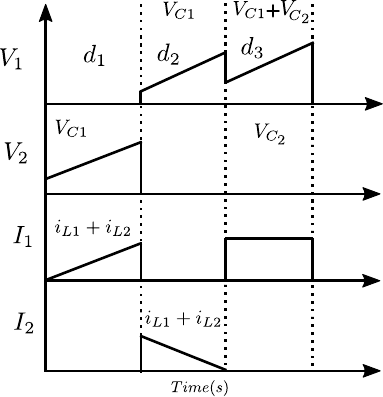}
\caption{Waveforms of $V_1$, $V_2$, $I_1$ and $I_2$}
\label{fig:Switches Voltages and Currents}
\end{figure}
\begin{align} 
<V_1> &= D_2*(V_{c1})+D_3*(V_{c1}+V_{c2})\\
<V_2> &= D_1*(V_{c1})-D_3(V{c2}) \\
<I_1> &= D_1*(i_{L1}+i_{L2}) \\
<I_2> &= D_2*(i_{L1}+i_{L2})
\end{align}

Therefore, the equivalent circuit would remain the same as that of the SEPIC.
\section{Circuit Averaging for a Non Ideal Cuk}
Fig. 10 shows a non-ideal Cuk converter operating in DCM with the switches separated
\begin{figure}[htbp!]
\centering
\includegraphics[scale=1]{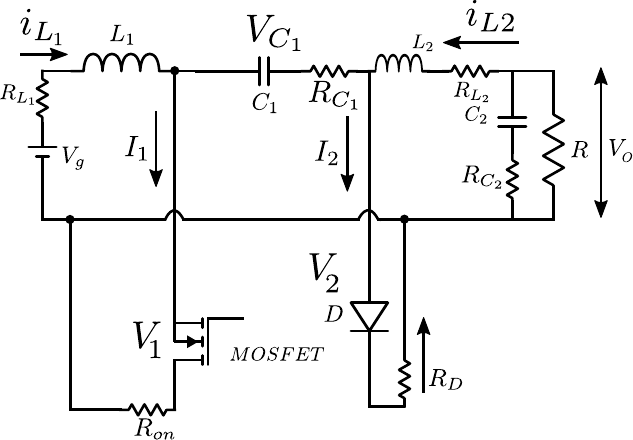}
\caption{Switch Separated}
\label{fig:Swicthes separated}
\end{figure}

On Averaging the voltages and currents across the switches

\begin{align} 
<V_1> &=((i_{L1}+i_{L2})R_{on1})D_1+(V_{c1}+V_d+
\end{align}
\begin{align*} 
R_d(i_{L1}+i_{L2}))D_2+D_3(V_{c1}+V_0) 
\end{align*}
\begin{align} 
<V_2> &=(V_{c1}- (i_{L1}+i_{L2})R_{on1})D_1-
\end{align}
\begin{align*} 
D_2(V_d+R_d(i_{L1}+i_{L2}))-D_3V_0 
\end{align*}
\begin{align} 
<I_1> &=(i_{L1}+i_{L2})D_1 \\
<I_2> &=(i_{L1}+i_{L2})D_2
\end{align}

It was observed that the equivalent circuit for (27) and (28) are similar to that of Fig. 4. Hence, the derived average model for two switch PWM DC-DC converter is generic and can be applied to any converter operating in DCM. 
The switch network is replaced by the equivalent circuit using CCM –DCM1 under average.lib in LTSpice.

\section{Specifications of the converters}

Assuming the converters’ operation in DCM, the specifications are selected. Table 1 and 2 show the specifications of Non ideal SEPIC and Cuk converters

\begin{table}[htbp] 
\center
\caption{Specifications of Non Ideal SEPIC}

\begin{tabular}{>{\flushleft}m{1.9in}     >{\flushleft}m{1in}}
\hline
 
\rule{0pt}{3ex}   \textbf{Parameters}  & \rule{0pt}{3ex}  \textbf{Value}   \\ \hline
\rule{0pt}{3ex} Input Voltage ($V_g$)
& \rule{0pt}{3ex} 62 V \\ \hline

\rule{0pt}{3ex} Output Voltage ($V_o$) & \rule{0pt}{3ex}  22 V  \\ \hline

\rule{0pt}{3ex} Output Resistance, R
 & \rule{0pt}{3ex}  52 $\Omega$  \\ \hline
 \rule{0pt}{3ex} Inductor, $L_1$
 & \rule{0pt}{3ex}  13 mH  \\ \hline
  \rule{0pt}{3ex} Inductor, $L_2$
 & \rule{0pt}{3ex}  166 $\mu$H  \\ \hline
 \rule{0pt}{3ex} Inductor ESR, $R_{L1}$
 & \rule{0pt}{3ex}  130m$\Omega$  \\ \hline
  \rule{0pt}{3ex} Inductor ESR, $R_{L2}$
 & \rule{0pt}{3ex}  110m$\Omega$  \\ \hline
   \rule{0pt}{3ex} MOSFET Resistance, $R_{on_1}$
 & \rule{0pt}{3ex}  31m$\Omega$  \\ \hline
    \rule{0pt}{3ex} Duty Cycle, D
 & \rule{0pt}{3ex}  0.2  \\ \hline
  \rule{0pt}{3ex} Capacitor, $C_1$
 & \rule{0pt}{3ex}  0.5 $\mu$H  \\ \hline
   \rule{0pt}{3ex} Capacitor, $C_2$
 & \rule{0pt}{3ex}  1000 $\mu$H  \\ \hline
   \rule{0pt}{3ex} Capacitor ESR, $R_{C1}$
 & \rule{0pt}{3ex}  270m$\Omega$  \\ \hline
    \rule{0pt}{3ex} Capacitor ESR, $R_{C2}$
 & \rule{0pt}{3ex}  110m$\Omega$  \\ \hline
     \rule{0pt}{3ex} Switching Frequency, $f_s$
 & \rule{0pt}{3ex}  50kHz  \\ \hline
      \rule{0pt}{3ex} Diode Drop, $V_d$
 & \rule{0pt}{3ex}  0.7V  \\ \hline
   \rule{0pt}{3ex} Diode Forward Resistance, $R_d$
 & \rule{0pt}{3ex}  0.12 $\Omega$  \\ \hline
  
\end{tabular}

\label{tab:tab1}
\end{table}
\begin{table}[htbp] 
\center
\caption{Specifications of Non Ideal Cuk}

\begin{tabular}{>{\flushleft}m{1.9in}     >{\flushleft}m{1in}}
\hline
 
\rule{0pt}{3ex}   \textbf{Parameters}  & \rule{0pt}{3ex}  \textbf{Value}   \\ \hline
\rule{0pt}{3ex} Input Voltage ($V_g$)
& \rule{0pt}{3ex} 25 V \\ \hline

\rule{0pt}{3ex} Output Voltage ($V_o$) & \rule{0pt}{3ex}  -21 V  \\ \hline

\rule{0pt}{3ex} Output Resistance, R
 & \rule{0pt}{3ex}  100 $\Omega$  \\ \hline
 \rule{0pt}{3ex} Inductor, $L_1$
 & \rule{0pt}{3ex}  1 mH \\ \hline
  \rule{0pt}{3ex} Inductor, $L_2$
 & \rule{0pt}{3ex}  1mH \\ \hline
 \rule{0pt}{3ex} Inductor ESR, $R_{L1}$
 & \rule{0pt}{3ex}  0.15$\Omega$  \\ \hline
  \rule{0pt}{3ex} Inductor ESR, $R_{L2}$
 & \rule{0pt}{3ex}  0.2$\Omega$  \\ \hline
   \rule{0pt}{3ex} MOSFET Resistance, $R_{on_1}$
 & \rule{0pt}{3ex}  31m$\Omega$  \\ \hline
    \rule{0pt}{3ex} Duty Cycle, D
 & \rule{0pt}{3ex}  0.42  \\ \hline
  \rule{0pt}{3ex} Capacitor, $C_1$
 & \rule{0pt}{3ex}  850 $\mu$H  \\ \hline
   \rule{0pt}{3ex} Capacitor, $C_2$
 & \rule{0pt}{3ex}  47 $\mu$H  \\ \hline
   \rule{0pt}{3ex} Capacitor ESR, $R_{C1}$
 & \rule{0pt}{3ex}  0.2$\Omega$  \\ \hline
    \rule{0pt}{3ex} Capacitor ESR, $R_{C2}$
 & \rule{0pt}{3ex}  0.3$\Omega$  \\ \hline
     \rule{0pt}{3ex} Switching Frequency, $f_s$
 & \rule{0pt}{3ex}  20kHz  \\ \hline
      \rule{0pt}{3ex} Diode Drop, $V_d$
 & \rule{0pt}{3ex}  0.75V  \\ \hline
   \rule{0pt}{3ex} Diode Forward Resistance, $R_d$
 & \rule{0pt}{3ex}  0.11$\Omega$  \\ \hline
  
\end{tabular}

\label{tab:tab1}
\end{table}

\section{Combined Model for CCM-DCM }
The advantages of using such model is (a) Simulation of CCM / DCM operation can be achieved in the same model (b) The decision is taken by the model and is made internal to the circuit
CCM/DCM 1 is an averaged block available under average.lib in LTSpice. A common equation satisfying CCM and DCM operation is shown below

For CCM and DCM operations, one of the governing equations is shown in [6] and (4). Where $\mu$ is the duty cycle in DCM operation.

\begin{align} 
<I_1> &= V_1/R_e \\
<V_1> &= ((1-\mu)/\mu)V_2
\end{align}
Where $\mu$ is the duty cycle in DCM operation.

Substituting, (30) in (4),
\begin{align} 
\mu &= \frac{V_2}{V_2+I_1R_e} \\
\mu &= \frac{1}{1+(R_eI_1)/V_1} \\
\mu &= D
\end{align}
(32) and (33) define D for the converter in CCM and DCM operations. 
Combining them,
\begin{align} 
\mu &= max(d,\frac{1}{1+(R_eI_1)/V_1})
\end{align}

It can be noted from [1], that $\mu_{DCM}$ $>$ $\mu_{CCM}$. The model uses two inputs viz,.(a) $L_{eq}$ = $L_1$ $L_2$/($L_1$+$L_2$) and (b)$f_s$

\section{Results}

Simulations were performed using LTSpice software package. The equivalent switch network was available as a built-in library under ‘average.lib’, CCM-DCM1. D was varied from 0.2 to 0.9 in steps of 0.01 and $V_0$ was analyzed. It was observed from Fig. 11 that $V_0$ and $i_{L1}$ increased with the increase in duty cycle. However, $i_{L2}$ decreased when the duty cycle increased which describes the working of a typical SEPIC.
\begin{figure}[htbp!]
\centering
\includegraphics[scale=0.4]{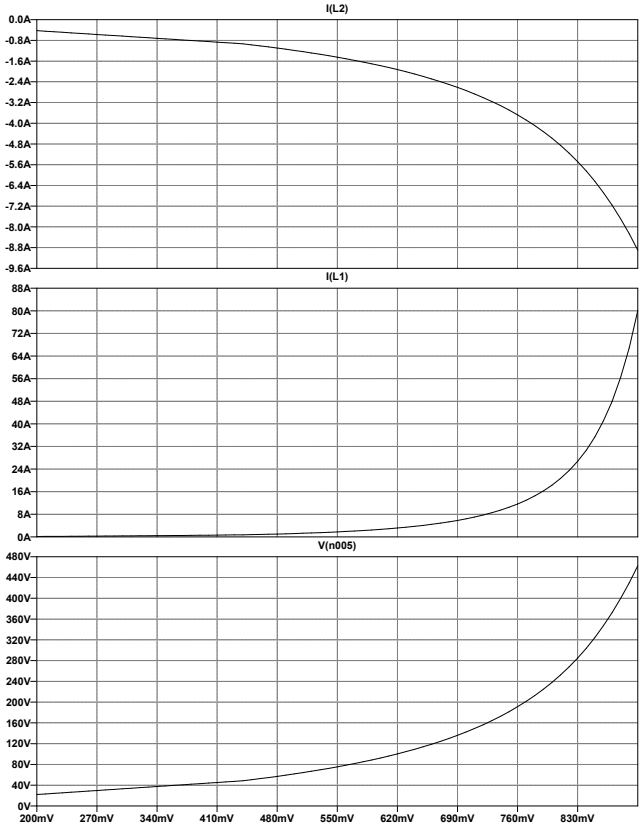}
\caption{$V_0$, $I_{L1}$ and $I_{L2}$ Vs. Time}
\label{fig:V0, IL1 & IL2 Vs. Time}
\end{figure}
Varying D, step changes in $R_{L1}$ and $R_{L2}$ were applied. $V_0$ and $i_{L1}$ were captured for the changes made. From Fig. 12 it was observed that highest value of $R_{L1}$ and lowest value of $R_{L2}$ showed maximum $V_0$.
\begin{figure}[htbp!]
\centering
\includegraphics[scale=0.38]{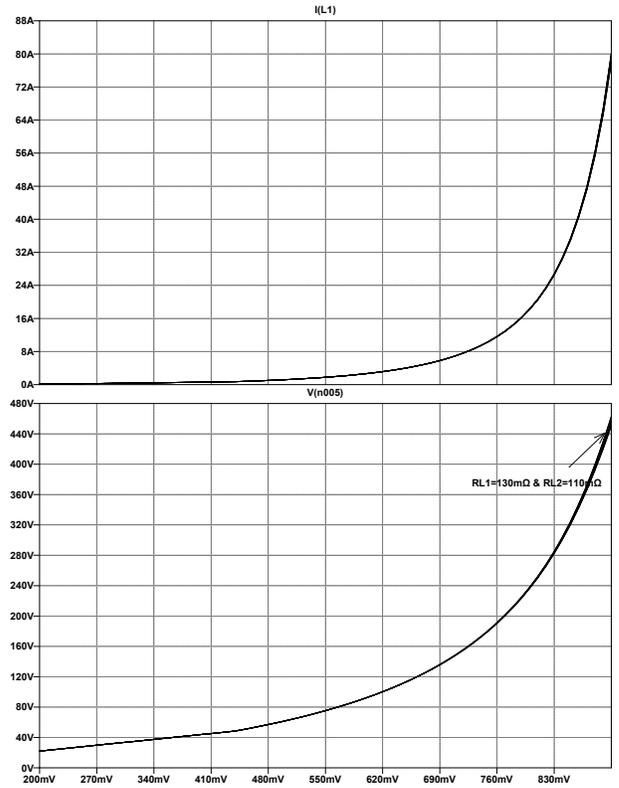}
\caption{$V_0$, $i_{L1}$ Vs. Time}
\label{fig:V0 and IL1 Vs. Time}
\end{figure}
Fig. 13 shows the bode plot of $G_{vd}$ for a fixed load R. It was observed from that the phase crossover frequency was around $1.814$ kHz, Gain margin around $5.254$ dB, gain cross over frequency around $76.834$ kHz, phase margin around $92.845^0$.
\begin{figure}[htbp!]
\centering
\includegraphics[scale=0.38]{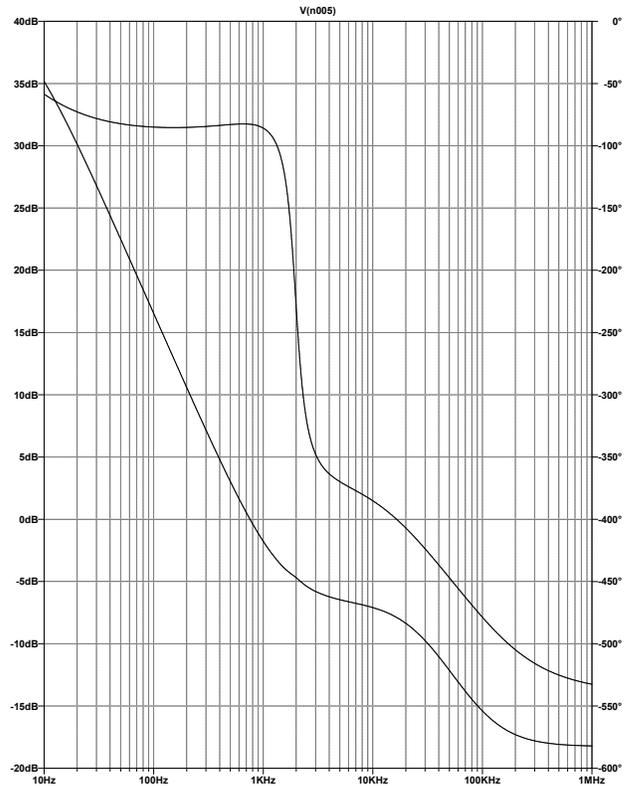}
\caption{Bode plot of $G_{vd}$}
\label{fig:Bode Plot of Gvd}
\end{figure}
\begin{figure}[htbp!]
\centering
\includegraphics[scale=0.38]{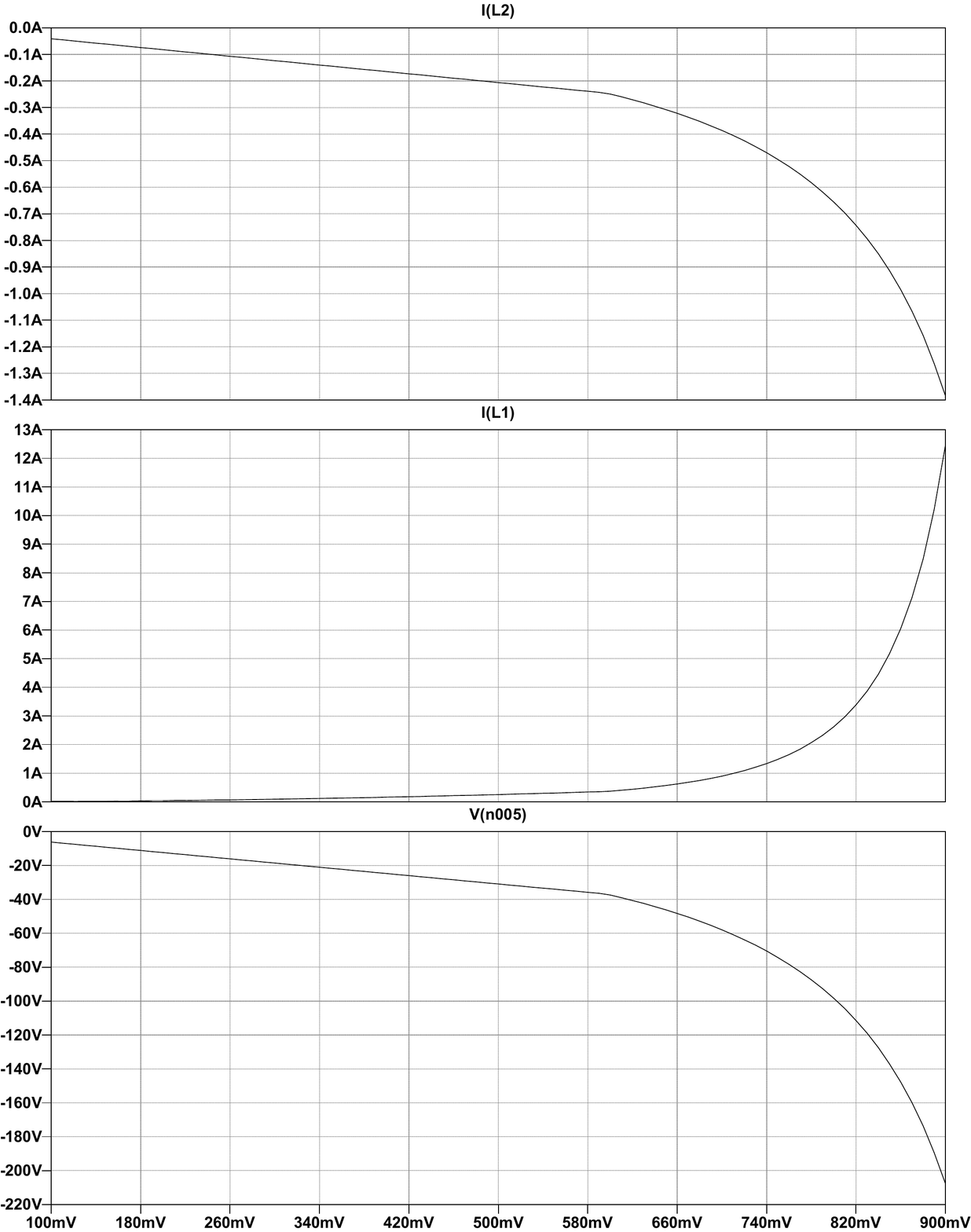}
\caption{$V_0$, $i_{L1}$, $i_{L2}$ Vs. Time for Cuk converter}
\label{fig:V0, iL1, iL2 Vs. Time for a Cuk converter}
\end{figure}
\begin{figure}[htbp!]
\centering
\includegraphics[scale=0.38]{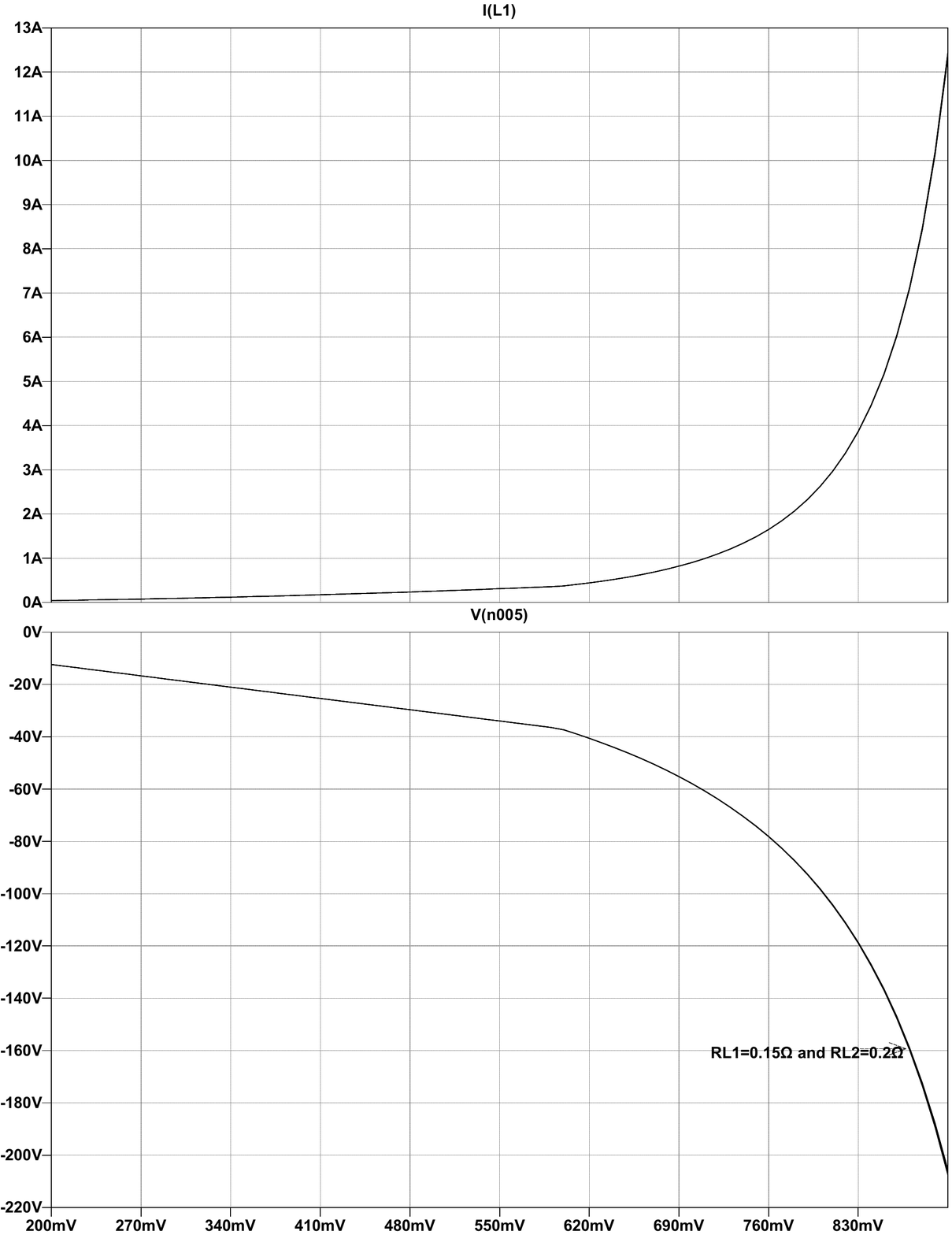}
\caption{$V_0$, $i_{L1}$ Vs. Time for Cuk converter}
\label{fig:V0, iL1 Vs. Time for a Cuk converter}
\end{figure}
\begin{figure}[htbp!]
\centering
\includegraphics[scale=0.38]{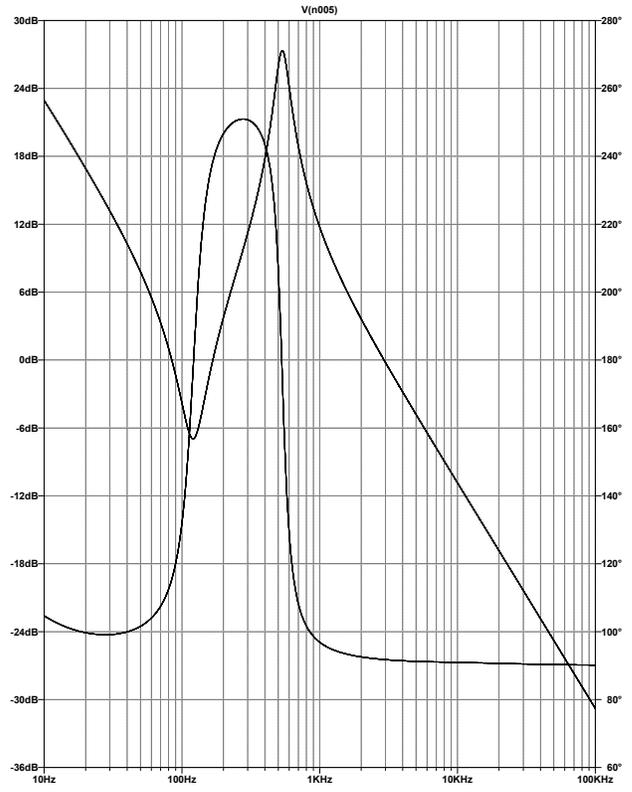}
\caption{$G_{vd}$ for Cuk converter}
\label{fig:Gvd for a Cuk converter}
\end{figure}
Fig.14 shows the nature of $V_0$, $i_{L1}$ and $i_{L2}$ for varying D proportional to the the control voltage, $V_c$ for a Cuk converter. It was observed that as D increased, $V_0$ and $i_{L2}$ decreased which defines the typical working of the Cuk converter.
Varying D, steps changes in $R_{L1}$ and $R_{L2}$ were applied. It can be observed from Fig. 15 that highest $R_{L1}$ and least $R_{L2}$ showed maximum $V_0$.
Fig. 16 shows the frequency response of $G_vd$ for a fixed load R. The gain margin was to be infinity, phase cross frequency of $81.982$ kHz and phase margin around $200.391^0$.
\section{Conclusion}
In this paper, the circuit averaging technique for fourth order converters like Cuk and SEPIC was carried out to obtain the frequency response for $G_{vd}$ using LTSpice simulation. This method can be generalized to find the response for any two switch DC-DC converters operating in CCM / DCM. This helps in developing an efficient feedback control design. Higher D produced higher $V_0$ in the converters. An appropriate controller to achieve sufficient gain margin and phase margin in closed loop operation and DCM analyses for isolated converters using CCM/DCM2 block are recommended.

\end{document}